\documentclass[prl,amsmath,amssymb,preprintnumbers,nofootinbib, reprint,
notitlepage
,longbibliography]{revtex4-1}
\pdfoutput=1
\usepackage[utf8]{inputenc} 
\usepackage{graphicx}
\usepackage{url}
\usepackage[bookmarks, pagebackref=false]{hyperref}
\usepackage[usenames,dvipsnames]{xcolor}
\usepackage{cleveref}

\definecolor{orange}{cmyk}{0,0.5,1,0}
\definecolor{graa}{rgb}{0.8,0.8,0.8}
\definecolor{blaa}{rgb}{0.2,0.2,0.6}

\definecolor{colA}{HTML}{c19277}
\definecolor{colB}{HTML}{e1bc91}
\definecolor{colD}{HTML}{62959c}

\hypersetup{
  colorlinks, 
  bookmarksopen, 
  bookmarksnumbered,
  citecolor=colA, 		
  linkcolor=colA,	
  urlcolor=colD,			
}
\usepackage{amsthm}
\usepackage{bm}
\usepackage{bbm}
\usepackage{pxfonts}

\usepackage{amsmath,amssymb,amsfonts}
\usepackage{color}
\usepackage{hyperref}
\usepackage[Symbolsmallscale]{upgreek}
\usepackage{amsmath}
\usepackage{amsfonts}
\usepackage{amssymb,dsfont}
\usepackage{graphicx}
\usepackage{amssymb}
\usepackage{bm}

\usepackage{placeins}
\usepackage{xspace}
\usepackage{cancel} 
\usepackage{slashed}
\usepackage{natbib}

\usepackage{tikz}
\usetikzlibrary{intersections}
\usetikzlibrary{arrows,decorations.pathmorphing,backgrounds,positioning,fit,petri}
\usetikzlibrary{decorations.markings}
\usetikzlibrary{calc}
\usetikzlibrary{intersections,through,hobby}
\usetikzlibrary{arrows.meta}

\pgfdeclarelayer{edgelayer}
\pgfdeclarelayer{nodelayer}
\pgfsetlayers{edgelayer,nodelayer,main}
\usepackage{adjustbox}
\newcommand{\lam}{\ensuremath{g}}

\usepackage[usenames,dvipsnames]{xcolor}
\usepackage[colorinlistoftodos,prependcaption,textsize=tiny]{todonotes}

\tikzstyle{none}=[inner sep=0mm]
\tikzstyle{dot}=[fill=black, draw=black, shape=circle,minimum size=8pt]
\tikzstyle{prop}=[-, draw=red, line width=2.5pt]
\tikzstyle{propsp}=[draw=blue, line width=2.5pt,
postaction={decorate,decoration={
    markings,
    mark=at position .55 with {\arrow[#1]{Latex[length=5mm, width=5mm]}}
  }}]
\tikzstyle{legsQ}=[double distance=6pt, draw=black, line width=2.5pt]
\tikzstyle{legsP}=[draw=black, line width=2.5pt]
\tikzstyle{legSP}=[draw=blue, line width=2.5pt,
postaction={decorate,decoration={
    markings,
    mark=at position .55 with {\arrow[#1]{Latex[length=5.5mm, width=4.5mm]}}
  }}]

\begin{document}

\title{
  \Large\boldmath Exact off-shell Sudakov form factor in $\mathcal{N}=4$ SYM}

\author{\sc A.V. Belitsky$^a$, L.V. Bork$^b$, A.F. Pikelner$^c$, V.A. Smirnov$^d$}
\affiliation{$^a$Department of Physics, Arizona State University, Tempe, AZ 85287-1504, USA}
\affiliation{$^b$Institute for Theoretical and Experimental Physics, 117218 Moscow, Russia}
\affiliation{The Center for Fundamental and Applied Research, 127030 Moscow, Russia}
\affiliation{$^c$Bogoliubov Laboratory of Theoretical Physics, Joint Institute for Nuclear Research, 141980~Dubna, Russia}
\affiliation{$^d$Skobeltsyn Institute of Nuclear Physics, Moscow State University 119992 Moscow, Russia}
\affiliation{Moscow Center for Fundamental and Applied Mathematics 119992 Moscow, Russia}

\begin{abstract}
We consider the Sudakov form factor in $\mathcal{N}=4$ SYM in the off-shell kinematical regime, which can be achieved by considering the theory on its Coulomb branch. We demonstrate that up two three loops both the infrared-divergent as well as the finite terms do exponentiate, with the coefficient accompanying $\log^2(m^2)$ determined by the octagon anomalous dimension $\Gamma_{oct}$. This behaviour is in strike contrast to previous conjectural accounts in the literature. Together with the finite terms we observe that up to three loops the logarithm of the Sudakov form factor is identical to twice the logarithm of the \emph{null octagon} $\mathbb{O}_0$, which was recently introduced within the context of integrability-based approaches to four point correlation functions with infinitely-large R-charges. The \emph{null octagon} $\mathbb{O}_0$ is known in a closed form for all values of the 't Hooft coupling constant and kinematical parameters. We conjecture that the relation between $\mathbb{O}_0$ and the off-shell Sudakov form factor will hold to all loop orders.
\end{abstract}

\maketitle

\section{Introduction}
\label{sec:intro}

Gauge theories are the cornerstones of the Standard Model of Particle Physics. Its unbroken gauge symmetries yield massless (Abelian) photons and (non-Abelian) gluons. Qualitatively, a highly energetic state, charged under corresponding gauge groups, emits vast amounts of these low-energy bosons as it propagates through the vacuum without experiencing any recoil. This implies that the original bare state is not what is measured by the detector, rather it is the one dressed by a cloud of soft bremsstrahlung. This compound is the new physical state of the theory. Quantitatively, all scattering amplitudes involving either the aforementioned bare state alone or it being accompanied by a finite number of gauge bosons decay exponentially fast $\exp(- \langle n \rangle/2)$ with the average number of soft-collinear gauge bosons $\langle n \rangle$ diverging double-logarithmically in an infrared  cutoff $m$ as $\langle n \rangle = \alpha \log^2 m$, $\alpha > 0$. This is the well-known infrared (IR) catastrophe. A finite result is then obtained provided one takes into account an infinite number of accompanying soft gauge bosons, i.e., for the dressed physical state. The precise cancellation mechanism is governed by the Kinoshita-Lee-Nauenberg theorem. As a result, the dependence on $m$ cancels out but a finite remainder is left, so one has to know the precise form of the accompanying coefficient $\alpha$. This can be done by studying the IR behavior of (virtual) quantum corrections to the scattering amplitude of $\ell$ bare states on an external source $\mathcal{O}$. The quantity in question is known as the form factor and its IR double-logarithmic limit as the Sudakov form factor \cite{Sudakov:1954sw},
\begin{align}
\label{Fdef}
F=
\langle 1, 2, \dots, \ell | \mathcal{O} |0 \rangle/\langle 1, 2, \dots, \ell | \mathcal{O} |0 \rangle_{\rm tree}
\, .
\end{align}
Apart from being of great interest in its own right, it encodes the IR structure of multi-particle scattering amplitudes ubiquitous to any high-energy scattering calculation.

For Abelian gauge theories, like QED, the Sudakov form factor is known to be one-loop exact, i.e., the average number of soft photon emissions, and thus $\alpha$, does not receive correction beyond the first loop order. Both off- \cite{Sudakov:1954sw} and on-shell \cite{Jackiw:1968zz} bare states were analyzed  and the difference in the corresponding values of $\alpha$ was found to be two, i.e., $\alpha_{\rm off} = 2 \alpha_{\rm on}$. The doubling is a consequence of an additional integration domain \cite{Fishbane:1971jz,Mueller:1981sg}, dubbed the ultra-soft, in loop momenta giving leading contributions on par with soft-collinear regions intrinsic to both.

In non-Abelian theories, such as QCD, the situation is far from being obvious. First, the non-commutativity of gauge bosons destroys the Poissonian nature of their emission,---they are no longer independent,--- and the coefficient $\alpha$ receives quantum effects to all orders in the Planck's constant. Second, existing literature \cite{Korchemsky:1988hd} suggests that QCD merely echoes the QED story and the off-/on-shell discrepancy is again the very same factor of two. The goal of this study is to demonstrate that this conclusion is precarious and the change requires the introduction of a new function of the coupling rather than just an overall multiplicative constant.

QCD is notoriously hard to solve, even in the quest for reaching sufficiently high orders of its perturbative series. So practitioners in the field often rely on a simpler QCD cousin, the four-dimensional maximally supersymmetric Yang-Mills theory, aka $\mathcal{N} = 4$ SYM. Both theories share similar properties at weak coupling. Folklore has it that for many observables the most complicated portions of QCD results coincide with complete contributions of the latter theory. This is known as the principle of maximal transcendentality \cite{Kotikov:2004er}. $\mathcal{N} = 4$ SYM is believed to be an integrable theory in the planar limit so one may hope to find closed-form expressions for matrix elements such as scattering amplitudes or form factors.

The on-shell form factors received a great deal of attention within $\mathcal{N} = 4$ SYM, starting from Ref.\ \cite{vanNeerven:1985ja} where $\ell = 2$ case was considered at second order of perturbation theory and culminating with recent studies that reached two-loop accuracy for four-leg amplitudes \cite{Guo:2021bym} and a staggering eight-loop order for a three-particle state \cite{Dixon:2022rse}. Moreover, form factors with more than two legs are amenable to integrability-based techniques \cite{Sever:2020jjx} deeply rooted in their duality to periodic null Wilson loops \cite{Maldacena:2010kp}. The IR exponent $\alpha$ of their soft limit is always driven by a universal function known as the cusp anomalous  dimension $\Gamma_{\rm cusp}$ \cite{Polyakov:1980ca}. The latter is known as a solution to a flux-tube integral equation \cite{Beisert:2006ez} at any value of the gauge coupling in planar limit. It is widely believed that $\Gamma_{\rm cusp}$ governs the IR behavior of all matrix elements in the theory such as scattering amplitudes or form factors in many kinematical regimes.

The off-shell case, on the other hand, received virtually no attention up until very recently, merely being indulged a discussion in passing, see, e.g., \cite{Drummond:2007aua}, and mirroring the QCD conjecture alluded to above. Recent results of Ref.\ \cite{Caron-Huot:2021usw} however suggest that the actual difference between on- and off-shell matrix elements in $\mathcal{N}=4$ SYM is far more involved than previously thought. Ref.\ \cite{Caron-Huot:2021usw} found that the IR behaviour of the four-gluon amplitude in $\mathcal{N}=4$ SYM on the Coulomb branch (i.e., theory with the spontaneously broken gauge symmetrty), which can be considered as the amplitude in the off-shell kinematics, is \emph{not} driven by $\Gamma_{\rm cusp}$, but rather by a complitely different function $\Gamma_{\rm oct}$. Two-loop computations of a five-leg amplitude in similar kinematics also point toward the conclusion that the IR asymptotic is controlled by $\Gamma_{\rm oct}$ \cite{Bork:2022vat}. This raises an immediate question: what is the true IR behaviour of the off-shell Sudakov form factor in $\mathcal{N}=4$ SYM and other gauge theories such as QCD in light of its paramount role in soft-gluon physics? 

The main practical aim of this paper is to report on a calculation of the two-leg off-shell Sudakov form factor $F$ to three loops in $\mathcal{N}=4$ SYM. Our findings can be summarised by the following concise formula for $\log F$:
\begin{align}\label{form1}
\log F &= -\frac{\Gamma_{\rm oct}(\lam)}{2}\log^2 \left( t \right)-D(\lam)+\mathcal{O}(m^2),
\end{align}
where the bra-state of the matrix element in its left-hand side depends on the outgoing particles' momenta $p_{1,2}$, obeying the off-shell condition $-p_{i}^2=m^2$. Momentum conservation reduces the dependence of $F$ only to $q=p_1+p_2$, which is the momentum incoming to the composite operator $\mathcal{O}$ such that $F$ depends only on the dimensionless variable $t\equiv m^2/Q^2$ with Euclidean $-q^2 = Q^2>0$. The asymptotic behavior of the Sudakov form factor in the limit $m^2 \to 0$ is determined by two functions of the 't Hooft coupling\footnote{We define the 't Hooft coupling constant as $g^2=g_{\scriptstyle\rm YM}^2N_c/(4 \pi)^2$.} $g$, $\Gamma_{\rm oct}(\lam)$ and $D(\lam)$, which are given by the following elementary functions:
\begin{align}\label{GamaOctDfunct}
\Gamma_{\rm oct}(\lam) & = \frac{2}{\pi^2}\log \left( \cosh \left(2\pi \lam\right)\right),\nonumber\\
D(\lam) & = \frac{1}{4}\log \left(\frac{\sinh( 4\pi \lam )}{4\pi \lam} \right).
\end{align}
Surprisingly not only the logarithmic term but also the finite function admits a closed-form expression in $g$ and coincide with the corresponding expressions of Refs.\ \cite{Belitsky:2019fan} obtained for a completely different object in $\mathcal{N} = 4$ SYM, which the authors of \cite{Caron-Huot:2021usw} conjectured to be dual to off-shell scattergin amplitudes.

\section{Techniques used}
\label{sec:tech}
\begin{figure}[h]
  \centering
  \begin{tabular}[t]{ccc}
    \adjustbox{valign=c}{\resizebox{2cm}{!}{\begin{tikzpicture}
	\begin{pgfonlayer}{nodelayer}
		\node [style=dot] (0) at (-4, -2) {};
		\node [style=dot] (1) at (4, -2) {};
		\node [style=dot] (2) at (0, 4) {};
		\node [style=none] (3) at (0, 6) {};
		\node [style=none] (4) at (-5, -4) {};
		\node [style=none] (5) at (5, -4) {};
	\end{pgfonlayer}
	\begin{pgfonlayer}{edgelayer}
		\draw [style=legsQ] (3.center) to (2);
		\draw [style=legsP] (0) to (4.center);
		\draw [style=legsP] (1) to (5.center);
		\draw [style=prop] (2) to (0);
		\draw [style=prop] (0) to (1);
		\draw [style=prop] (1) to (2);
	\end{pgfonlayer}
\end{tikzpicture}}}
    &
      \adjustbox{valign=c}{\resizebox{2cm}{!}{\begin{tikzpicture}
	\begin{pgfonlayer}{nodelayer}
		\node [style=dot] (0) at (-4, -2) {};
		\node [style=dot] (1) at (4, -2) {};
		\node [style=dot] (2) at (0, 4) {};
		\node [style=none] (3) at (0, 6) {};
		\node [style=none] (4) at (-5, -4) {};
		\node [style=none] (5) at (5, -4) {};
		\node [style=dot] (6) at (-2, 1) {};
		\node [style=dot] (7) at (2, 1) {};
	\end{pgfonlayer}
	\begin{pgfonlayer}{edgelayer}
		\draw [style=legsQ] (3.center) to (2);
		\draw [style=legsP] (0) to (4.center);
		\draw [style=legsP] (1) to (5.center);
		\draw [style=prop] (0) to (6);
		\draw [style=prop] (6) to (2);
		\draw [style=prop] (2) to (7);
		\draw [style=prop] (7) to (1);
		\draw [style=prop] (0) to (1);
		\draw [style=prop] (6) to (7);
	\end{pgfonlayer}
\end{tikzpicture}}}
    &
      \adjustbox{valign=c}{\resizebox{2cm}{!}{\begin{tikzpicture}
	\begin{pgfonlayer}{nodelayer}
		\node [style=dot] (0) at (-4, -2) {};
		\node [style=dot] (1) at (4, -2) {};
		\node [style=dot] (2) at (0, 4) {};
		\node [style=none] (3) at (0, 6) {};
		\node [style=none] (4) at (-5, -4) {};
		\node [style=none] (5) at (5, -4) {};
		\node [style=dot] (6) at (-2, 1) {};
		\node [style=dot] (7) at (2, 1) {};
		\node [style=none] (8) at (-0.5, -0.25) {};
		\node [style=none] (9) at (0.5, 0.25) {};
	\end{pgfonlayer}
	\begin{pgfonlayer}{edgelayer}
		\draw [style=legsQ] (3.center) to (2);
		\draw [style=legsP] (0) to (4.center);
		\draw [style=legsP] (1) to (5.center);
		\draw [style=prop] (0) to (6);
		\draw [style=prop] (6) to (2);
		\draw [style=prop] (2) to (7);
		\draw [style=prop] (7) to (1);
		\draw [style=prop] (6) to (1);
		\draw [style=prop] (0) to (8.center);
		\draw [style=prop] (7) to (9.center);
	\end{pgfonlayer}
\end{tikzpicture}}}\\
    $T_{1,1}$ & $T_{2,1}$ & $T_{2,2}$
  \end{tabular}
  \caption{Scalar integrals contributing to $F$ at one and two loop order. Double external line corresponds to the momentum $q=p_1+p_2$ carried by the operator. Thin external lines corresponds to the particles' momenta $p_{1,2}$. All internal lines are massless.}
  \label{fig:ints12l}
\end{figure}
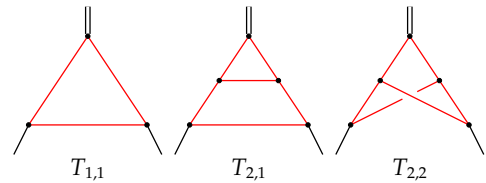

\begin{figure*}[t]
  \centering
  \begin{tabular}[t]{cccccccc}
    \adjustbox{valign=c}{\resizebox{2cm}{!}{\begin{tikzpicture}
	\begin{pgfonlayer}{nodelayer}
		\node [style=dot] (0) at (-4, -2) {};
		\node [style=dot] (1) at (4, -2) {};
		\node [style=dot] (2) at (0, 5) {};
		\node [style=none] (3) at (0, 7) {};
		\node [style=none] (4) at (-5, -4) {};
		\node [style=none] (5) at (5, -4) {};
		\node [style=dot] (6) at (-2, 2) {};
		\node [style=dot] (7) at (2, 2) {};
		\node [style=dot] (8) at (-3, 0) {};
		\node [style=dot] (9) at (3, 0) {};
	\end{pgfonlayer}
	\begin{pgfonlayer}{edgelayer}
		\draw [style=legsQ] (3.center) to (2);
		\draw [style=legsP] (0) to (4.center);
		\draw [style=legsP] (1) to (5.center);
		\draw [style=prop] (0) to (1);
		\draw [style=prop] (0) to (8);
		\draw [style=prop] (8) to (6);
		\draw [style=prop] (6) to (2);
		\draw [style=prop] (2) to (7);
		\draw [style=prop] (7) to (9);
		\draw [style=prop] (9) to (1);
		\draw [style=prop] (6) to (7);
		\draw [style=prop] (8) to (9);
	\end{pgfonlayer}
\end{tikzpicture}}}
    &
      \adjustbox{valign=c}{\resizebox{2cm}{!}{\begin{tikzpicture}
	\begin{pgfonlayer}{nodelayer}
		\node [style=dot] (0) at (-4, -2) {};
		\node [style=dot] (1) at (4, -2) {};
		\node [style=dot] (2) at (0, 5) {};
		\node [style=none] (3) at (0, 7) {};
		\node [style=none] (4) at (-5, -4) {};
		\node [style=none] (5) at (5, -4) {};
		\node [style=dot] (6) at (-2, 2) {};
		\node [style=dot] (7) at (2, 2) {};
		\node [style=dot] (8) at (0, 2) {};
		\node [style=dot] (9) at (0, -2) {};
    \node [style=none,anchor=south west] (pa) at (1.5, 3.5) {\Huge $p_a$};
    \node [style=none,anchor=north east] (pb) at (4, -3) {\Huge $p_b$};    
	\end{pgfonlayer}
	\begin{pgfonlayer}{edgelayer}
		\draw [style=legsQ] (3.center) to (2);
		\draw [style=legsP] (0) to (4.center);
		\draw [style=prop] (6) to (2);
		\draw [style=propsp] (2) to (7);
		\draw [style=prop] (6) to (0);
		\draw [style=prop] (7) to (1);
		\draw [style=prop] (6) to (8);
		\draw [style=prop] (8) to (7);
		\draw [style=prop] (0) to (9);
		\draw [style=prop] (9) to (1);
		\draw [style=prop] (8) to (9);
		\draw [style=legSP] (5.center) to (1);
	\end{pgfonlayer}
\end{tikzpicture}}}
    &
      \adjustbox{valign=c}{\resizebox{2cm}{!}{\begin{tikzpicture}
	\begin{pgfonlayer}{nodelayer}
		\node [style=dot] (0) at (-4, -2) {};
		\node [style=dot] (1) at (4, -2) {};
		\node [style=dot] (2) at (0, 5) {};
		\node [style=none] (3) at (0, 7) {};
		\node [style=none] (4) at (-5, -4) {};
		\node [style=none] (5) at (5, -4) {};
		\node [style=dot] (6) at (-2, 2) {};
		\node [style=dot] (7) at (2, 2) {};
		\node [style=dot] (8) at (-3, 0) {};
		\node [style=dot] (9) at (3, 0) {};
		\node [style=none] (10) at (0.25, 1.25) {};
		\node [style=none] (11) at (-0.25, 1) {};
    \node [style=none,anchor=south east] (pa) at (-1.5, 3.5) {\Huge $p_a$};
    \node [style=none,anchor=south east] (pb) at (-4, -1) {\Huge $p_b$};    
	\end{pgfonlayer}
	\begin{pgfonlayer}{edgelayer}
		\draw [style=legsQ] (3.center) to (2);
		\draw [style=legsP] (0) to (4.center);
		\draw [style=legsP] (1) to (5.center);
		\draw [style=prop] (0) to (1);
		\draw [style=propsp] (0) to (8);
		\draw [style=prop] (8) to (6);
		\draw [style=prop] (2) to (7);
		\draw [style=prop] (7) to (9);
		\draw [style=prop] (9) to (1);
		\draw [style=propsp] (2) to (6);
		\draw [style=prop] (6) to (9);
		\draw [style=prop] (7) to (10.center);
		\draw [style=prop] (8) to (11.center);
	\end{pgfonlayer}
\end{tikzpicture}}}
    &
      \adjustbox{valign=c}{\resizebox{2cm}{!}{\begin{tikzpicture}
	\begin{pgfonlayer}{nodelayer}
		\node [style=dot] (0) at (-4, -2) {};
		\node [style=dot] (1) at (4, -2) {};
		\node [style=dot] (2) at (0, 5) {};
		\node [style=none] (3) at (0, 7) {};
		\node [style=none] (4) at (-5, -4) {};
		\node [style=none] (5) at (5, -4) {};
		\node [style=dot] (6) at (-2, 2) {};
		\node [style=dot] (7) at (2, 2) {};
		\node [style=dot] (8) at (-3, 0) {};
		\node [style=dot] (9) at (3, 0) {};
		\node [style=none] (10) at (-0.75, -0.75) {};
		\node [style=none] (11) at (0.5, -1) {};
    \node [style=none,anchor=south west] (pa) at (3, 1) {\Huge $p_a$};
    \node [style=none,anchor=north east] (pb) at (4, -3) {\Huge $p_b$};    
	\end{pgfonlayer}
	\begin{pgfonlayer}{edgelayer}
		\draw [style=legsQ] (3.center) to (2);
		\draw [style=legsP] (0) to (4.center);
		\draw [style=prop] (0) to (8);
		\draw [style=prop] (8) to (6);
		\draw [style=prop] (6) to (2);
		\draw [style=prop] (2) to (7);
		\draw [style=propsp] (7) to (9);
		\draw [style=prop] (9) to (1);
		\draw [style=prop] (6) to (7);
		\draw [style=legSP] (5.center) to (1);
		\draw [style=prop] (0) to (9);
		\draw [style=prop] (8) to (10.center);
		\draw [style=prop] (1) to (11.center);
	\end{pgfonlayer}
\end{tikzpicture}}}
    &
      \adjustbox{valign=c}{\resizebox{2cm}{!}{\begin{tikzpicture}
	\begin{pgfonlayer}{nodelayer}
		\node [style=dot] (0) at (-4, -2) {};
		\node [style=dot] (1) at (4, -2) {};
		\node [style=dot] (2) at (0, 5) {};
		\node [style=none] (3) at (0, 7) {};
		\node [style=none] (4) at (-5, -4) {};
		\node [style=none] (5) at (5, -4) {};
		\node [style=dot] (6) at (-2, 2) {};
		\node [style=dot] (7) at (2, 2) {};
		\node [style=dot] (8) at (-3, 0) {};
		\node [style=dot] (9) at (3, 0) {};
		\node [style=none] (10) at (1, 1.25) {};
		\node [style=none] (11) at (0.5, 0.75) {};
		\node [style=none] (12) at (-1, -0.25) {};
		\node [style=none] (13) at (-1.75, -0.75) {};
    \node [style=none,anchor=south west] (pa) at (3, 1) {\Huge $p_a$};
    \node [style=none,anchor=north east] (pb) at (4, -3) {\Huge $p_b$};    
	\end{pgfonlayer}
	\begin{pgfonlayer}{edgelayer}
		\draw [style=legsQ] (3.center) to (2);
		\draw [style=legsP] (0) to (4.center);
		\draw [style=prop] (0) to (8);
		\draw [style=prop] (8) to (6);
		\draw [style=prop] (6) to (2);
		\draw [style=prop] (2) to (7);
		\draw [style=propsp] (7) to (9);
		\draw [style=prop] (9) to (1);
		\draw [style=prop] (8) to (1);
		\draw [style=prop] (6) to (9);
		\draw [style=prop] (7) to (10.center);
		\draw [style=prop] (11.center) to (12.center);
		\draw [style=prop] (13.center) to (0);
		\draw [style=legSP] (5.center) to (1);
	\end{pgfonlayer}
\end{tikzpicture}}}
    &
      \adjustbox{valign=c}{\resizebox{2cm}{!}{\begin{tikzpicture}
	\begin{pgfonlayer}{nodelayer}
		\node [style=dot] (0) at (-4, -2) {};
		\node [style=dot] (1) at (4, -2) {};
		\node [style=dot] (2) at (0, 5) {};
		\node [style=none] (3) at (0, 7) {};
		\node [style=none] (4) at (-5, -4) {};
		\node [style=none] (5) at (5, -4) {};
		\node [style=dot] (6) at (-2, 2) {};
		\node [style=dot] (7) at (2, 2) {};
		\node [style=dot] (8) at (-3, 0) {};
		\node [style=dot] (9) at (3, 0) {};
		\node [style=none] (10) at (1, 1.25) {};
		\node [style=none] (11) at (0.5, 0.75) {};
		\node [style=none] (12) at (-1, -0.25) {};
		\node [style=none] (13) at (-1.75, -0.75) {};
    \node [style=none,anchor=south east] (pa) at (-3, 1) {\Huge $p_a$};
    \node [style=none,anchor=south west] (pb) at (4, -1) {\Huge $p_b$};    
	\end{pgfonlayer}
	\begin{pgfonlayer}{edgelayer}
		\draw [style=legsQ] (3.center) to (2);
		\draw [style=legsP] (0) to (4.center);
		\draw [style=legsP] (1) to (5.center);
		\draw [style=prop] (0) to (8);
		\draw [style=prop] (6) to (2);
		\draw [style=prop] (2) to (7);
		\draw [style=prop] (7) to (9);
		\draw [style=propsp] (9) to (1);
		\draw [style=prop] (8) to (1);
		\draw [style=prop] (6) to (9);
		\draw [style=prop] (7) to (10.center);
		\draw [style=prop] (11.center) to (12.center);
		\draw [style=prop] (13.center) to (0);
		\draw [style=propsp] (6) to (8);
	\end{pgfonlayer}
\end{tikzpicture}}}
    &
      \adjustbox{valign=c}{\resizebox{2cm}{!}{\begin{tikzpicture}
	\begin{pgfonlayer}{nodelayer}
		\node [style=dot] (0) at (-4, -2) {};
		\node [style=dot] (1) at (4, -2) {};
		\node [style=dot] (2) at (0, 5) {};
		\node [style=none] (3) at (0, 7) {};
		\node [style=none] (4) at (-5, -4) {};
		\node [style=none] (5) at (5, -4) {};
		\node [style=dot] (8) at (-3, 0) {};
		\node [style=dot] (9) at (3, 0) {};
		\node [style=dot] (10) at (0, 1) {};
	\end{pgfonlayer}
	\begin{pgfonlayer}{edgelayer}
		\draw [style=legsQ] (3.center) to (2);
		\draw [style=legsP] (0) to (4.center);
		\draw [style=legsP] (1) to (5.center);
		\draw [style=prop] (0) to (1);
		\draw [style=prop] (0) to (8);
		\draw [style=prop] (9) to (1);
		\draw [style=prop] (8) to (2);
		\draw [style=prop] (2) to (9);
		\draw [style=prop] (8) to (10);
		\draw [style=prop] (10) to (9);
		\draw [style=prop] (2) to (10);
	\end{pgfonlayer}
\end{tikzpicture}}}
    &
      \adjustbox{valign=c}{\resizebox{2cm}{!}{\begin{tikzpicture}
	\begin{pgfonlayer}{nodelayer}
		\node [style=dot] (0) at (-4, -2) {};
		\node [style=dot] (1) at (4, -2) {};
		\node [style=dot] (2) at (0, 5) {};
		\node [style=none] (3) at (0, 7) {};
		\node [style=none] (4) at (-5, -4) {};
		\node [style=none] (5) at (5, -4) {};
		\node [style=dot] (8) at (-3, 0) {};
		\node [style=dot] (9) at (3, 0) {};
		\node [style=dot] (10) at (0, -2) {};
		\node [style=none] (11) at (0, 0.5) {};
		\node [style=none] (12) at (0, -0.5) {};
	\end{pgfonlayer}
	\begin{pgfonlayer}{edgelayer}
		\draw [style=legsQ] (3.center) to (2);
		\draw [style=legsP] (0) to (4.center);
		\draw [style=legsP] (1) to (5.center);
		\draw [style=prop] (0) to (8);
		\draw [style=prop] (9) to (1);
		\draw [style=prop] (8) to (2);
		\draw [style=prop] (2) to (9);
		\draw [style=prop] (8) to (9);
		\draw [style=prop] (0) to (10);
		\draw [style=prop] (10) to (1);
		\draw [style=prop] (2) to (11.center);
		\draw [style=prop] (12.center) to (10);
	\end{pgfonlayer}
\end{tikzpicture}}}
    \\
    $T_{3,1}$ & $T_{3,2}$ & $T_{3,3}$ & $T_{3,4}$& $T_{3,5}$& $T_{3,6}$& $T_{3,7}$& $T_{3,8}$
  \end{tabular}
  \caption{Scalar integrals contributing to $F$ at three loop level. Arrows and labels $p_a$, $p_b$ on the lines correspond to the presence of numerator $(p_a+p_b)^2$. All other notations are identical to those of fig.\ref{fig:ints12l}.}
  \label{fig:ints3l}
\end{figure*}
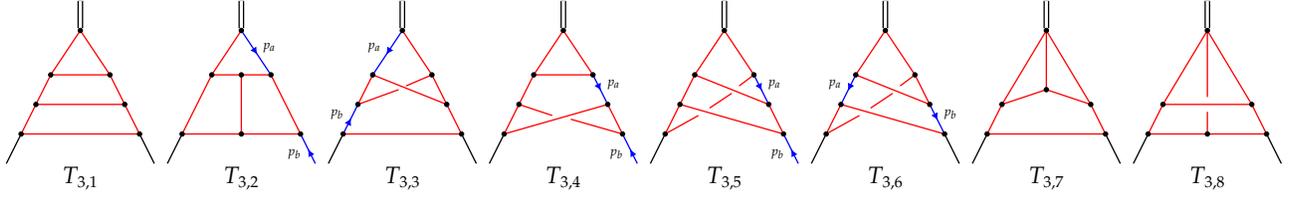

The starting point of our analysis was Eq.\ (\ref{Fdef}) for $\ell=2$, with $\mathcal{O}$ being the lowest component of the stress-tensor supermultiplet \cite{Brandhuber:2010ad,Bork:2010wf,Brandhuber:2011tv,Bork:2011cj}, in the off-shell Euclidean kinematical regime introduced at the end the previous section. All states propagating in quantum loops are strictly massless. To do this in a gauge-invariant and self-consistent manner we relied on the approach advocated by Refs.\ \cite{HennGiggs1,Caron-Huot:2021usw}, which is based on the observation that amplitude integrands on the Coulomb branch of the $\mathcal{N}=4$ SYM are equivalent to the ones in the maximally supersymmetric theory with unbroken gauge symmetry but in higher dimension, i.e., $D>4$ \cite{Caron-Huot:2021usw,HenrietteCoulomb,HennGiggs1}. The extra dimensional components of the massless $D$-dimensional momenta are then interpreted, from the four-dimensional perspective, as masses \cite{Caron-Huot:2021usw,HennGiggs1}. Further we adopted yet another observation which states that loop integrands of the two-leg form factor, similarly to four-leg amplitudes considered in \cite{Caron-Huot:2021usw}, are identical in all even dimensions $D\leq10$, at least for the first few orders of perturbation theory \cite{Gehrmann:2011xn}. 

The last remark allowed us to recycle Feynman graphs of Ref.\ \cite{Gehrmann:2011xn} defining the four-dimensional on-shell Sudakov form factor for our $D$-dimensional integrands with the same topologies and accompanying coefficients. Then we chose $D>4$ components of all momenta in the integrand in such a manner that all internal lines of scalar integrals are strictly massless in $D=4$, while all external lines become massive (or, equivalently, obey the off-shell conditions $-p_i^2=m^2 \to 0$). Integrations over loop momenta are done strictly in $D=4$. This off-shell form factor can then be interpreted as a form factor of a pair of W-bosons in the small mass limit similarly to the point of view of \cite{Caron-Huot:2021usw} for the off-shell four-leg scattering amplitude. 

Up to three loops, the off-shell Sudakov form factor is given by the 
perturbative expansion
\begin{equation}\label{LogMoffshell1}
F = 1+ \lam^2 \,F_1 + \lam^4 \, F_2 + \lam^6 \,F_3 + \mathcal{O}(\lam^8),\\
\end{equation}
with one and to loop corrections given by the following scalar integrals:
\begin{equation}\label{LogMoffshell2}
F_1 = 2 Q^2~T_{1,1},\quad F_2= Q^4 \left(4T_{2,1}+T_{2,2}\right),
\end{equation}
see Fig.\ \ref{fig:ints12l}, and the three-loop term being
\begin{align}\label{LogMoffshell3}
F_3 = Q^4&\left( Q^2~8 T_{3,1}-2T_{3,2}+4T_{3,3}+4T_{3,4}\right.\nonumber\\
& \left. \quad -4T_{3,5}
-4T_{3,6}-4T_{3,7}+2T_{3,8}\right),
\end{align}
see Fig.\ \ref{fig:ints3l}.
At one and two loops, all integrals can be expressed in terms of $l$-loop triangle ladder function $\Phi^{(l)}(x,y)$ with equal arguments $\Phi_l\equiv \Phi^{(l)}(t,t)$ \cite{Usyukina:1993ch}:
\begin{equation}\label{12loopIntegralsPlanarAndNPlanar}
  (-Q^2) T_{1,1}=\Phi_1,\quad (-Q^2)^2 T_{2,1}=\Phi_2,\quad (-Q^2)^2 T_{2,2}=\Phi_1^2.
\end{equation}
The small-$t$ expansion of $\Phi_{1,2}$ admits the following form:
\begin{align}\label{BoxSmallMassExp1}
\Phi_1 & =\log^2(t)+2\zeta_2 + \mathcal{O}(m^2),\nonumber\\
\Phi_2 & =\frac{\log^4(t)}{4}+3\zeta_2\log^2(t)+\frac{21\zeta_4}{2} + \mathcal{O}(m^2).
\end{align}
We would like to remind the reader that in contrast to the case of dimensional regularization there is no analog of $\epsilon \times 1/\epsilon$-interference between different order in 't Hooft coupling, and the relations (\ref{BoxSmallMassExp1}) are sufficient to completely determine $F_1$ and $F_2$ up to $O(m^2)$ terms.

The three-loop scalar integrals $T_{3,i}$ from Fig.~\ref{fig:ints3l}  are more involved compared to their one- and two-loop counterparts. These perfectly fit into the set of auxiliary integrals studied in Ref.\ \cite{Pikelner:2021goo}. We used the package \texttt{LiteRed} \cite{Lee:2012cn} for their reduction to a set of master integrals calculated in \cite{Pikelner:2021goo} making use of the method of differential equations~\cite{Kotikov:1990kg,Gehrmann:1999as,Henn:2013pwa}. The so-determined integrals were in turn expressed in terms of HPL's of the single argument $t$ of the weight not grater then $6$. After that they were expanded at small values of $t$ using the \texttt{HPL} package \cite{Maitre:2005uu}.

To partially cross-check our findings, we calculated the IR divergent, i.e., $\log^2(t)$, part making use of the strategy of regions \cite{Beneke:1997zp} (see also \cite{Smirnov:2002pj,Smirnov:2012gma}), reformulated in the language of the Feynman-parameter representation in \cite{Smirnov:1999bza}. We relied on their algorithmic determination with the help of the package {\tt asy} \cite{Pak:2010pt,Jantzen:2012mw} (also available with the {\tt FIESTA5} distribution package \cite{Smirnov:2021rhf}), which is based on the geometry of polytopes associated with Symanzik polynomials defining corresponding integrands. For each of the integrals involved, the strategy of regions yielded scaleless parametric integrals which were evaluated by using their Mellin-Barnes representation (see, e.g., Chapter~5 of \cite{Smirnov:2012gma}). We found a perfect agreement between these two approaches. Details regarding these computations will be published elsewhere. 

Expanding our three-loop results in powers of $t$ we found that logarithmic and constant parts of all $T_{3,i}$ integrals can be cast as the Davydychev-Usyukina functions
\begin{align}\label{TtoDRelation}
  (-Q^2)^3 T_{3,1} & = \Phi_3 + \mathcal{O}(m^2),\nonumber\\
  (-Q^2)^2 T_{3,2} & = \Phi_3 + \mathcal{O}(m^2),\nonumber\\
  (-Q^2)^2(T_{3,3}-T_{3,5}) & = \frac{1}{2}\left(\Phi_3-\Phi_1\Phi_2\right) + \mathcal{O}(m^2),\nonumber\\
  (-Q^2)^2(T_{3,4}-T_{3,6}) & = -\Phi_1\Phi_2 + \mathcal{O}(m^2),\nonumber\\
  (-Q^2)^2 T_{3,7} & = \Phi_3 + \mathcal{O}(m^2),\nonumber\\
  (-Q^2)^2 T_{3,8} & = \Phi_1\Phi_2 + \mathcal{O}(m^2),
\end{align}
where $\Phi_3$ develops the expansion as $t \to 0$ \begin{align}\label{BoxSmallMassExp2}
  \Phi_3 & =\frac{1}{36}\log^6(t)+\frac{5\zeta_2}{6}\log^4(t)
+\frac{35\zeta_4}{2}\log^2(t)\nonumber\\
&+\frac{155\zeta_6}{4}+\mathcal{O}(m^2).
\end{align}
Expanding $\log F$ in powers of $\lam$ we have found that, up to the three-loop order, $\log F$ is equal to:
\begin{equation}\label{LogM2octagon}
\log F \left(t,\lam\right) = -\frac{\Gamma_{\rm oct}(\lam)}{2}\log^2 \left(t\right)-D(\lam)+\mathcal{O}(m^2),
\end{equation}
with:
\begin{align}
\Gamma_{\rm oct}(\lam) & = 4 \lam^2 - 16 \zeta_2 \lam^4 + 256\zeta_4 \lam^6+\ldots,\nonumber\\
D(\lam) & = 4 \zeta_2 \lam^2 - 32 \zeta_4 \lam^4+\frac{1024\zeta_6}{3} \lam^6+\ldots.
\end{align}
This is exactly the logarithm of the \emph{null octagon} \cite{Coronado:2018cxj,Belitsky:2019fan} $\mathbb{O}_0(z,\bar{z})$ multiplied by $2$:
\begin{equation}
\log\mathbb{O}_0(z,\bar{z})=-\frac{\Gamma_{\rm oct}(\lam)}{4}\log^2\left(\frac{\bar z}{z}\right)
-\frac{\lam^2}{2}\log(z\bar z)-\frac{D(\lam)}{2},
\end{equation}
with $z \bar{z}=1,~\bar{z}=\sqrt{t}$. The functions of the 't Hooft coupling $\Gamma_{\rm oct}(\lam)$ and $D(\lam)$ are given to all orders of pertubation theory by the closed formulas (\ref{GamaOctDfunct}). We conjecture that this relation holds for all loops as well:
\begin{equation}\label{LogM2conjcture}
\log F = 2\log \mathbb{O}_0 + O(m^2).
\end{equation}

\section{Discussion and conclusion}
\label{sec:disc-concl}
We observe that the off-shell Sudakov form factor in $\mathcal{N}=4$ SYM reveals an intriguing and unbeknown to date structure. As was pointed out in the introduction, there was a conjecture in the literature  \cite{ConformalProperties4point}, for an all-order evolution equation that $F$ is anticipated to obey, namely,
\begin{eqnarray}\label{EvolEq1}
\frac{\partial \log F}{\partial \log m^2}=-\Gamma_{\rm cusp}(\lam) 
\log m^2 + \Gamma_{\rm col}(\lam) ,
\end{eqnarray}
where \cite{Beisert:2006ez}
\begin{eqnarray}
\Gamma_{\rm cusp}(\lam)=4\lam^2 - 8\zeta_2\lam^4 + 88 \zeta_4 \lam^6 +\ldots.
\end{eqnarray}
We conclude that this equation is valid only at the one-loop level and should be replaced with
\begin{eqnarray}\label{EvolEq3}
\frac{\partial \log F}{\partial \log m^2}
=
-\Gamma_{\rm oct}(\lam) \log m^2 .
\end{eqnarray}
There are two obvious differences, (i) the leading IR function is not $\Gamma_{\rm cusp}$, which is thought of as an ultimate IR exponent of all gauge theories, and (ii) the so-called collinear anomalous dimension $\Gamma_{\rm col}$ is absent in the off-shell kinematics, at least to three-loop order. Based on this, we believe that the same disparity persists between the two regimes in QCD as well. 

Prior to our current analysis there were earlier studies of form factors on the Coulomb branch where, however, the choice of scalar vacuum expectation values was done in a way such that all external states were massless but a virtual particle ``framing'' Feynman graphs was massive \cite{HennGiggs1,HennGiggs2,Henn:2011by}. In this case, the evolution equation was found to coincide with the one in the massless case (\ref{EvolEq1}), albeit with $\Gamma_{\rm cusp} \to \Gamma_{\rm cusp}/2$. Thus, we observe a very subtle, anomalous effect of the non-commutativity of $p_i^2\rightarrow 0$ and $\epsilon \rightarrow 0$ limits. We can expect that the situation in QCD will be similar. We will address these questions in upcoming publications in full detail.

Another mysterious relation we would like to unravel is how the off-shellness relates to the flux-tube origin of the IR exponents $\Gamma_{\rm cusp}$ and $\Gamma_{\rm oct}$ in $\mathcal{N}=4$ SYM.  It turns out that both of them can be obtained from a single deformed flux-tube integral equation \cite{Basso:2020xts}, which combines the two describing $\Gamma_{\rm cusp}$ \cite{Beisert:2006ez} and $\Gamma_{\rm oct}$ \cite{Belitsky:2019fan} separately.

Relations (\ref{TtoDRelation}) and (\ref{12loopIntegralsPlanarAndNPlanar}) between integrals also deserve a dedicated study. They can be thought of as a manifestation of the dual conformal symmetry of form factors which was anticipated for quite some time \cite{Bork:2014eqa, Bianchi:2018rrj}. 

\begin{acknowledgments}
L.B.\ is grateful to A.I.\ Onishchenko for useful discussions and to A.V.\
Bednyakov, N.B.\ Muzhichkov and E.S.\ Sozinov for collaboration at early stages
of the project. The work of A.B.\ was supported by the U.S. National Science
Foundation under the grant No.\ PHY-2207138. The work of L.B.\ is supported by
the Foundation for the Advancement of Theoretical Physics and Mathematics
"BASIS". The work of A.P.\ is supported by Russian Science Foundation under
grant 20-12-00205. The work of V.S.\ was supported by the Ministry of Education
and Science of the Russian Federation as part of the program of the Moscow
Center for Fundamental and Applied Mathematics under the Agreement No.
075-15-2019-1621.
\end{acknowledgments}

\bibliography{sudakovFF3l}
\end{document}